\documentclass[conference]{IEEEtran}
\IEEEoverridecommandlockouts
\usepackage{cite}
\usepackage{amsmath,amssymb,amsfonts}
\usepackage{algorithmic}
\usepackage{graphicx}
\usepackage{textcomp}
\usepackage{xcolor}
\usepackage{subcaption}
\usepackage{authblk}
\usepackage{hyperref}

\def\BibTeX{{\rm B\kern-.05em{\sc i\kern-.025em b}\kern-.08em
    T\kern-.1667em\lower.7ex\hbox{E}\kern-.125emX}}

\begin{document}

\title{Federated Phish Bowl: LSTM-Based Decentralized Phishing Email Detection}

\author{Yuwei Sun, Ng Chong, and Hideya Ochiai
\thanks{Yuwei Sun and Hideya Ochiai are with the Graduate School of Information Science and Technology, University of Tokyo, Tokyo, Japan (e-mail: ywsun@g.ecc.u-tokyo.ac.jp, ochiai@elab.ic.i.u-tokyo.ac.jp). Ng Chong is with United Nations University, Tokyo, Japan (e-mail: ngstc@unu.edu).}}

\maketitle

\begin{abstract}
With increasingly more sophisticated phishing campaigns in recent years, phishing emails lure people using more legitimate-looking personal contexts. To tackle this problem, instead of traditional heuristics-based algorithms, more adaptive detection systems such as natural language processing (NLP)-powered approaches are essential to  understanding phishing text representations. Nevertheless, concerns surrounding the collection of phishing data that might cover confidential information hinder the effectiveness of model learning. We propose a decentralized phishing email detection framework called Federated Phish Bowl (FedPB) which facilitates collaborative phishing detection with privacy. In particular, we devise a knowledge-sharing mechanism with federated learning (FL). Using LTSM for phishing detection, the framework adapts by sharing a global word embedding matrix across the clients, with each client running its local model with Non-IID data. We collected the most recent phishing samples to study the effectiveness of the proposed method using different client numbers and data distributions. The results show that FedPB can attain a competitive performance with a centralized phishing detector, with generality to various cases of FL retaining a prediction accuracy of 83\%.   
\end{abstract}

\begin{IEEEkeywords}
data privacy, federated learning, long short-term memory, multi-party computation, phishing email detection
\end{IEEEkeywords}

\section{Introduction}
The sharp uptick of phishing emails across the globe has exacerbated continued risks to personal data privacy and security in recent years. A recent report shows that phishing attacks soar 220\% during the COVID-19 peak \cite{f5}. A phishing email usually adopts legitimate-look contexts to deceive users and steal sensitive information such as credit card numbers, bank accounts, and passwords. Phishing emails target a broad range of fields with highly crafted text data based on social engineering and users' personal online experiences. Despite existing detection mechanisms, phishing emails continue to slip past organizations’ defenses. Notably, classical approaches such as signature-based detection can not work proactively due to ever-changing texts of phishing messages embedded with various types of hooks.

On the other hand, recent advancements in deep learning (DL) have made text mining of large numbers of different types of phishing emails possible. Knowledge acquisition and sentiment analysis based on natural language processing (NLP) offer a practical solution to the learning representation of text data concerning the tone, grammatical coherence, emotion, and so forth, significantly reducing human efforts in feature engineering. For this reason, it is considered that phishing emails’ common features can be extracted and learned by a deep neural network (DNN), such as the long short-term memory (LSTM) model for phishing detection. 

Despite this, training a DNN model usually needs vast data from both the phishing and legitimate categories. However,  legitimate emails containing sensitive personal information are difficult to collect in the real world. This difficulty renders typical centralized learning less adequate for tackling the task. In particular, data privacy concerns can refrain people from participating. A learned model on under-sampled training data can result in incorrect classification when encountering new emails unseen before. The challenge is how to learn a phishing detection model using distributed and skewed emails. It appears to be more reasonable to leverage a decentralized learning architecture to improve the system's adaptability to new samples by sharing model updates among clients. 

The main contributions of this paper are:

(1) There is a notable scarcity of publicly available phishing data for research purposes. Moreover, many such data sets are dated which may not reflect the more contemporary tactics pursued by bad actors. For this reason, we investigated the most recent phishing emails reported from different groups of users (Section \ref{datacollection}). 

(2) We applied a federated learning framework tailored to automate the detection of phishing emails called Federated Phish Bowl (FedPB) (Section \ref{fedpb}). 

(3) Extensive experiments were performed to evaluate the effectiveness of FedPB for automated phishing detection by varying the client number and the skewness of client training samples (Section \ref{experiments}).

The remainder of this paper is structured as follows. Section 2 presents the related work of phishing email detection based on machine learning. Section 3 presents essential definitions and assumptions. Section 4 demonstrates the technical underpinnings of the proposed method. Section 5 presents extensive empirical evaluations. Section 6 concludes the paper and gives out future directions.

\begin{table*}[!t]
  \centering
  \caption{ML Methodologies for Phishing Email Detection}\label{table1}
  \begin{tabular}{llll}
    \textbf{WORK} & \textbf{YEAR} & \textbf{MODEL TOPOLOGY} & \textbf{METHODOLOGY}\\
    \hline \\
    Gutierrez et al.\cite{gutierrez} & 2018 & Centralized & Random under-sampling boost\\
    Unnithan et al.\cite{unnithan} & 2018& Centralized & Deep neural networks\\ 
    Nguyen et al.\cite{nguyen} & 2018&Centralized & LSTM with an attention mechanism\\
    Smadi et al.\cite{smadi} & 2018&Centralized & Reinforcement learning\\
    Sahingoz et al.\cite{sahingoz} &2019& Centralized & Seven machine learning methods\\
    Fang et al.\cite{zhang} & 2019&Centralized & Recurrent convolutional neural networks\\
    Alhogail and Alsabih\cite{alhogail} & 2021&Centralized & Graph convolutional networks\\
    Thapa et al.\cite{thapa}& 2021 & Decentralized & Fine tuning-based methods 
\end{tabular}
\end{table*}

\section{Related Work}
Many methods have been proposed to safeguard email users from phishing, including traditional inspection and ML-based methods. In particular, traditional methods usually involve analysis of email formats, the integrity of email senders, and other attached meta information. This detection method is usually based on blacklists, heuristics, and visual analytics, which are neither feasible nor adaptive to real-life ever-changing phishing emails. On the other hand, ML-based methods offer promising solutions for large-scale knowledge acquisition from email lexical contents \ref{table1}. For instance, Sahingoz et al.\cite{sahingoz} demonstrated a real-time anti-phishing system and compared its performance with seven different ML methods. They suggested that the random forest had the best performance for phishing detection. Gutierrez et al.\cite{gutierrez} presented a random under-sampling boost (Reboots)-based method to build a retrainable system adaptive to newly observed samples. Moreover, deep neural networks (DNNs) \cite{dl} such as convolutional neural networks (CNNs) and recurrent neural networks (RNNs) have been adopted to improve detection by efficient text mining, alleviating efforts in feature engineering of domain experts. For example, Nguyen et al.\cite{nguyen} proposed a hierarchical attentive long short-term memory (LSTM)-based detection method that models the email bodies at the word level and the sentence level while leveraging a supervised attention mechanism. Furthermore, Smadi et al.\cite{smadi} presented reinforcement learning-based detection to reflect changes in newly explored behaviors, thus detecting zero-day phishing attacks. Fang et al.\cite{zhang} demonstrated the THEMIS where emails were modeled at the email header and body and the character and word level simultaneously. They verified its effectiveness by evaluating an unbalanced dataset of phishing and legitimate emails.

Though the aforementioned centralized methods provide a solution to security analysis of email contents, data processing on a collection of emails might violate the privacy of personal information. It necessitates an architecture that brings the gap between a centralized ML model and distributed training sources. However, there are not many studies of the decentralized phishing email detection. Thapa et al.\cite{thapa} presented a decentralized method of fine-tuning pre-trained language models for each participating client. However, the result showed that with the continual increasing of the client number, the model performance could greatly degrade. Moreover, the sharing of a large language model for every learning round could increase the communication cost regarding the model size, instead, we employing the global word embedding for local model fine-tuning of each client.

\section{Preliminaries}

\subsection{Phishing Email Analysis using Natural Language Processing}
\label{datacollection}
We introduce a new set of phishing emails collected recently from the Microsoft 365 anti-phishing protection \cite{365}. Notably, we extracted and decoded the emails that are labeled as high-confidence phishing by anti-phishing policies. In detail, phishing emails usually contain various types of information such as images, URLs, attachments, main texts, and so on. We mainly focus on the sequence data of main texts using natural language processing (NLP) to acquire meaningful knowledge in corpora. Moreover, we extract information from both the header and body of the email. We extract the subject of an email from the header and the main texts from the body, using the beautiful soup \cite{bs} to parse the embedded HTML pages in the email. 

\begin{figure}[!t]
\centerline{\includegraphics[width=\linewidth]{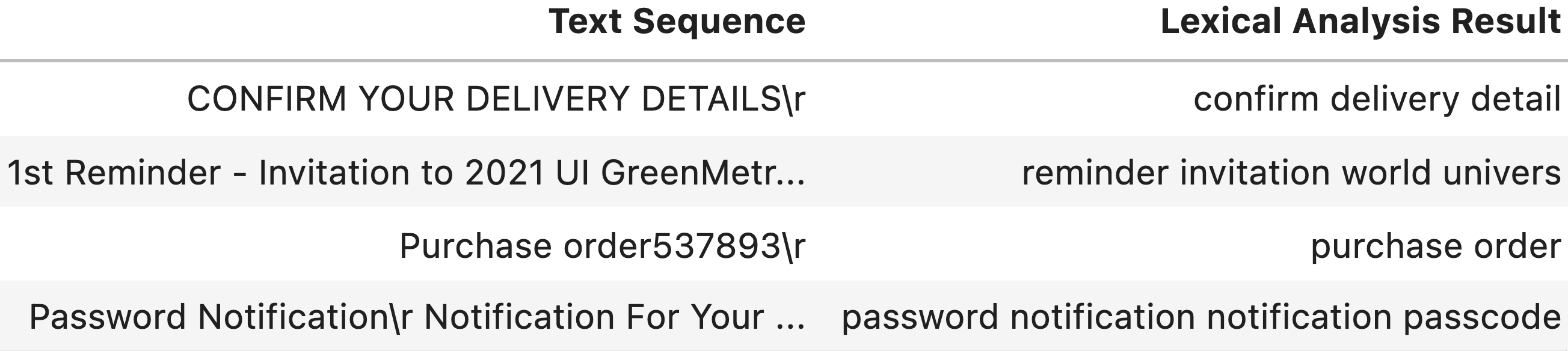}}
\caption{Cleaned and standardized email samples with irrelevant information removed by the NLP pipeline.}
\label{lexical}
\end{figure}

Furthermore, we concatenate the extracted subject and body texts and assign weights to important words by removing irrelevant information such as prepositions based on the following pipeline: (1) remove characters that are not letters such as numbers and punctuation marks, (2) convert the characters to lowercase, (3) extract tokens from the input string by splitting it into small chunks of words, (4) convert words with various tenses and plurals into their base or dictionary forms based on lemmatization, (5) remove any stop words such as “the”, “a”, and “is” and the words with less than two characters in the texts, (6) combine these cleaned and standardized tokens back into a continual string (Fig. \ref{lexical}). 

\subsection{Federated Learning}
There are usually two topologies to process distributed client emails based on machine learning, i.e., centralized learning and decentralized learning. Centralized learning leverages high-performance computing (HPC) to perform large-scale training on a collection of email samples. In contrast, decentralized learning leverages distributed model training on devices such as personal computers and smartphones. Federated learning (FL) is one of the decentralized learning approaches, which allows on-device model training without disclosing training data and achieves a better model by sharing trained models among clients. 

Suppose that there are $K$ clients and the parameter server (PS) in FL. Let $L_t^{(k)}$ be the local model of client$k$ and $G_t$ be the global model of the PS at round $t$. Every round, the PS randomly selects a small subset of $K_{selected}$ clients to perform local model training. Then, for each selected client $k\in K_{selected}$, the model is trained on its local dataset and the model update $L^{(k)}_t-G_t$ is sent to the PS. After the PS receives local updates from these clients, a model aggregation function such as the FedAvg \cite{fl} is adopted to update the global model based on local updates. The FedAvg algorithm which computes the weighted averaging of local updates can be formulated by the following
$G_{t+1} = G_t + \sum_{k\in K_{selected}}\frac{n_k}{n_{K_{selected}}}(L^{(k)}_t - G_t)$ where $n_k$ is the number of samples in client $k$'s local dataset and $n_{K_{selected}}$ is the number of total samples of all selected clients.

\section{Federated Phish Bowl}
\label{fedpb}

\begin{figure}[!t]
\centerline{\includegraphics[width=\linewidth]{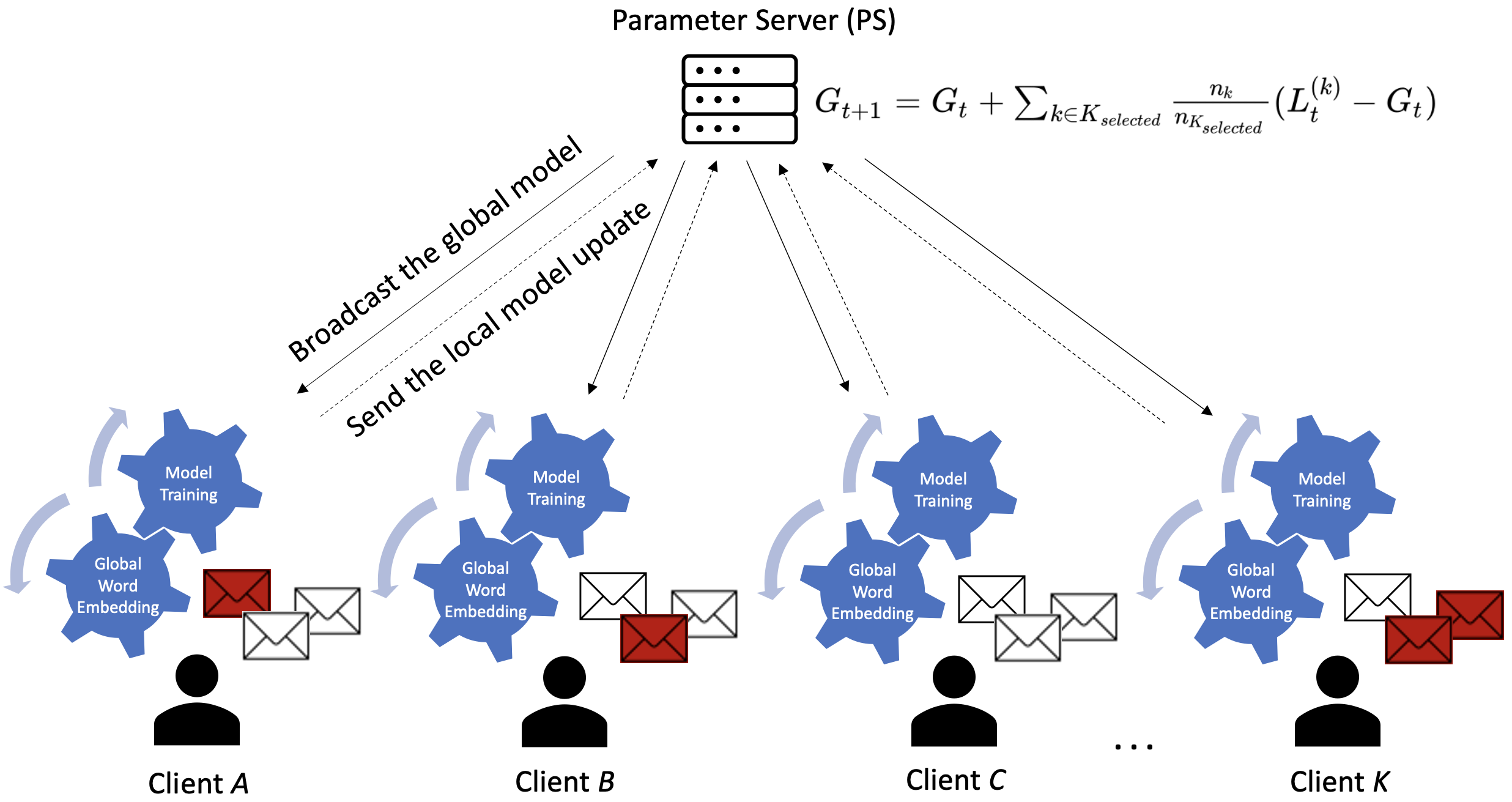}}
\caption{FedPB for the decentralized phishing email detection.}
\label{scheme}
\end{figure}

We propose the Federated Phish Bowl (FedPB), a decentralized phishing email detection system that allows adaptive knowledge representation and transfer of common phishing features among different clients (Fig. \ref{scheme}). We introduce the two building blocks of FedPB in the following sections, i.e., global word embedding and phishing email detection with bidirectional long short-term memory. 

\subsection{Feature Representation with Global Word Embedding}
Before processing email contents with neural networks, it necessitates representing the strings with numerical feature vectors. One approach to achieve this goal is to apply an embedding layer to learn a mapping from the input strings to vector representations for the phishing email detection task. However, training multiple embedding layers for different clients in FL could degrade model performance after the model aggregation in the PS. Moreover, though a large language model might be adaptive enough to retain the representation ability after the aggregation, communicating by such a large model can greatly increase the communication cost. 

On the other hand, the feature representations of phishing emails can also be transferred from other NLP tasks. In particular, we propose the global word embedding method by adopting a pre-learned word embedding in the PS and distributing the embedding to clients at the beginning of learning. This method can improve the adaptability of model learning and eliminate the need for each client in FL to train an individual embedding layer. Furthermore, this strategy relies on a word embedding called GloVe \cite{glove} for obtaining the feature representations of different words. The GloVe model is trained on a word-to-word co-occurrence matrix which tabulates how frequently words co-occur with one another in a given corpus. Notably, we employ the word embedding learned on six billion tokens from Wikipedia 2014 and Gigaword 5, which converts each input word to a 100-item feature vector.

\subsection{Phishing Email Detection with Bidirectional Long Short-Term Memory}

\begin{figure*}[!t]
  \centering
  \includegraphics[width=0.8\linewidth]{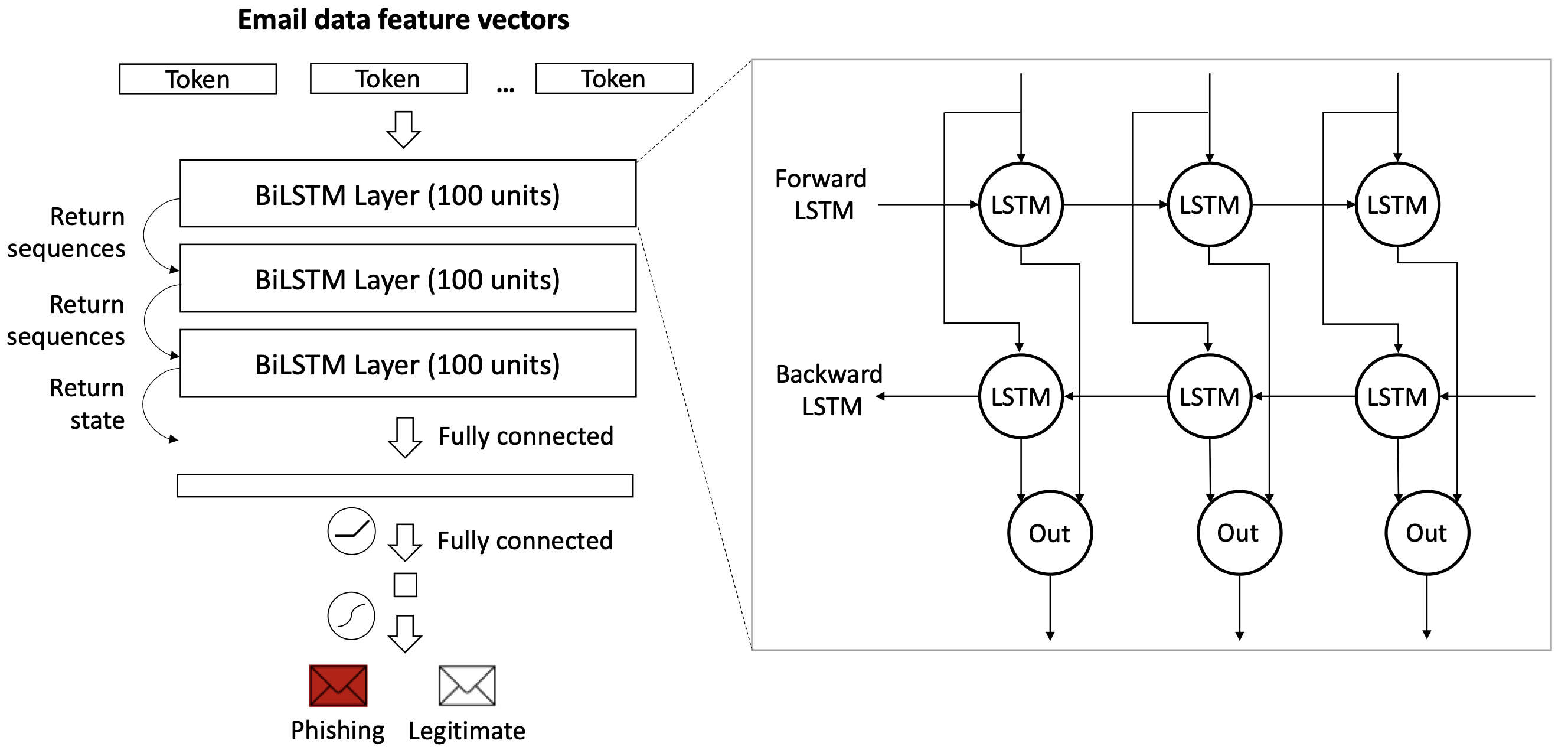}
\caption{The architecture of the learning model based on BiLSTM.}
\label{lstm}
\end{figure*}

Recurrent neural networks (RNNs) automatically extract hidden features from the input sequence data and classify these features in a high-dimensional space. To tackle the phishing email detection problem, we adopt an RNNs model called bidirectional long short-term memory (BiLSTM) \cite{bidirectional}, which processes the input sequence from positive and negative directions, thus improving its learning performance (Fig. \ref{lstm}). LSTM cells form the BiLSTM layer, which can remember long-term word associations leveraging the forget gate, the input gate, and the output gate. Finally, an output of the BiLSTM layer concatenates the computation results from the two directions.

In particular, we employ a five-layer neural network model consisting of three BiLSTM layers and two fully connected layers as the local task model of clients' model training. In detail, the input layer of the neural networks has a total of 200 time-steps, with each step having 100 embedded features from the result of global word embedding. Then, it is followed by three BiLSTM layers with 100 memory units each, where the first two layers output the state at each time step and send the sequence to the next layer. In contrast, the last BiLSTM layer returns the cell state for the last input time step and sends it to the fully connected layer, where a fully connected layer with 200 neurons is applied using as an activation function the ReLU. Finally, the output layer has a single neuron that predicts the type of an email, using the Sigmoid as an activation function. In addition, legitimate emails are assigned a label of 0, and phishing emails are assigned a label of 1 to train the model.

Furthermore, the learning process of the FedPB is as follows: (1) the parameter server (PS) initializes the global model of the BiLSTM neural networks, (2) the PS sends the global model and the global word embedding matrix to all clients, (3) the PS randomly selects a subset of clients to conduct local model training every round, (4) a selected client updates its model based on features represented by the word embedding matrix of local data, (5) the local update is sent back to the PS to update the current global model. As such, by sharing a global word embedding matrix and local model updates, the systems can achieve a better and better classifier for phishing email detection without the need for clients to reveal sensitive email contents. In addition, to facilitate the learning process, two identically structured global models can work simultaneously on the PS, where one model is for the training, and the other is for the inference. Then, for every several learning rounds, the inference model will be replaced by the training model.

\section{Experiments}
\label{experiments}
\subsection{Settings}
We collected the most recent high-confidence phishing emails quarantined by the Microsoft 365 anti-phishing protection \cite{365}, consisting of a total of 678 high-confidence phishing emails from the last three months. Then, using the aforementioned NLP-based lexical analysis and the global word embedding, we extracted essential text information from these phishing emails and transformed them into feature vectors. Moreover, to regulate the input sequence, we set 200 as the maximum feature vector length, a feature vector with a length greater than 200 being truncated. Whereas a vector with a length smaller than 200 was post-padded with zeros. In addition, any feature vector with a length smaller than 10 was removed. Finally, after removing all the under-length emails and duplicates, a total of 594 phishing email samples were retained for the dataset. To balance data in the dataset, we adopted 594 random legitimate email samples from the Enron dataset \cite{enron}. We obtained a dataset consisting of 1188 emails, each of which was represented with a 200-item feature vector. The dataset was separated into the training set and the test set with a ratio of 4:1.

\subsection{Numerical Results}

\begin{figure}
  \centering
  \includegraphics[width=0.86\linewidth]{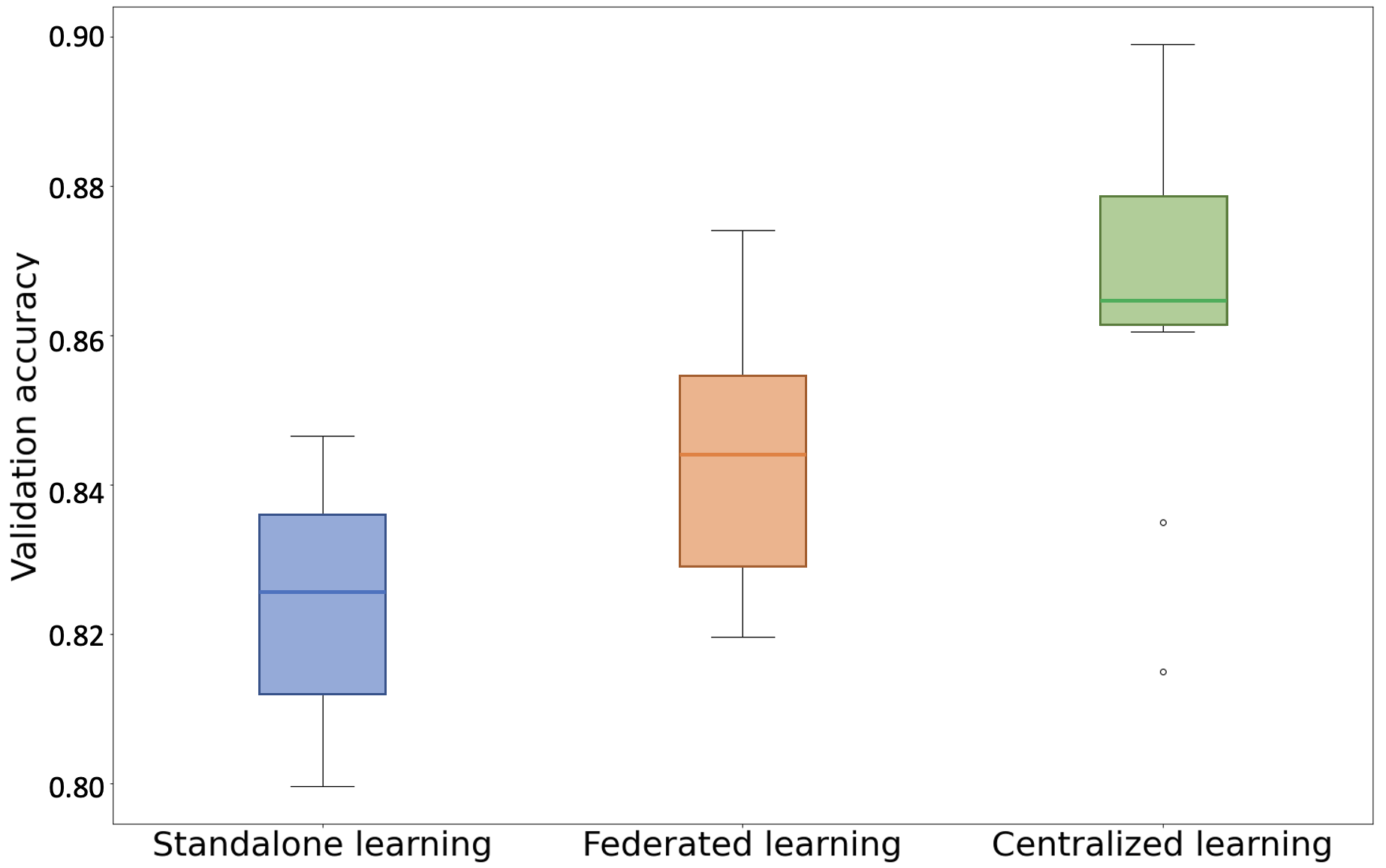}  
  \caption{Test accuracy of the three different methods for phishing email detection.}
  \label{result}
\end{figure}

\subsubsection{Detecting with Different Numbers of Clients}
\label{number}

\begin{figure}
  \centering
  \includegraphics[width=0.85\linewidth]{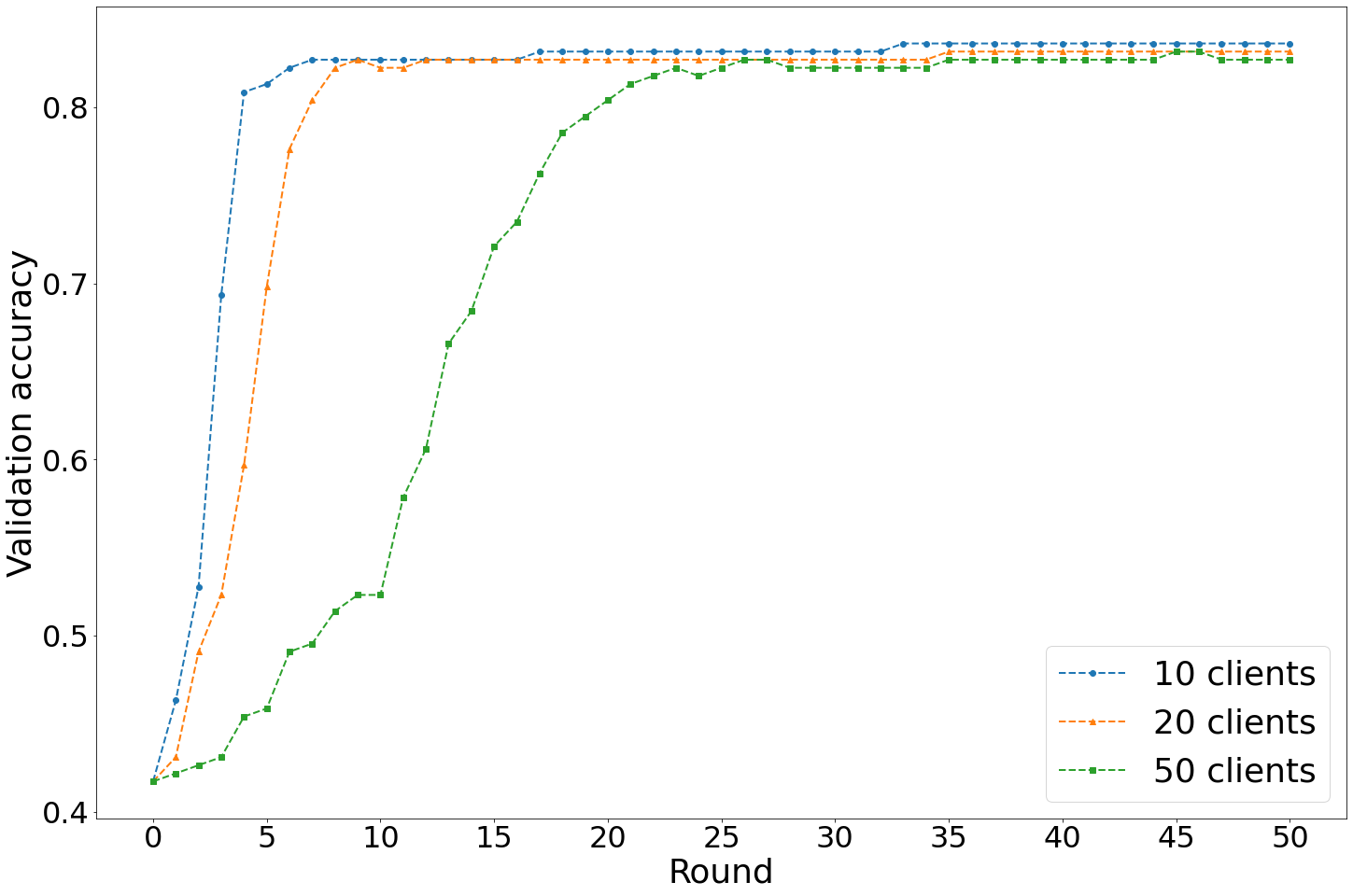}  
  \caption{Test accuracy of the global model when applying different numbers of clients in the FedPB.}
  \label{clientnum}
\end{figure} 

\begin{figure*}
\centering
    \begin{subfigure}[b]{0.33\textwidth}            
            \includegraphics[width=\textwidth]{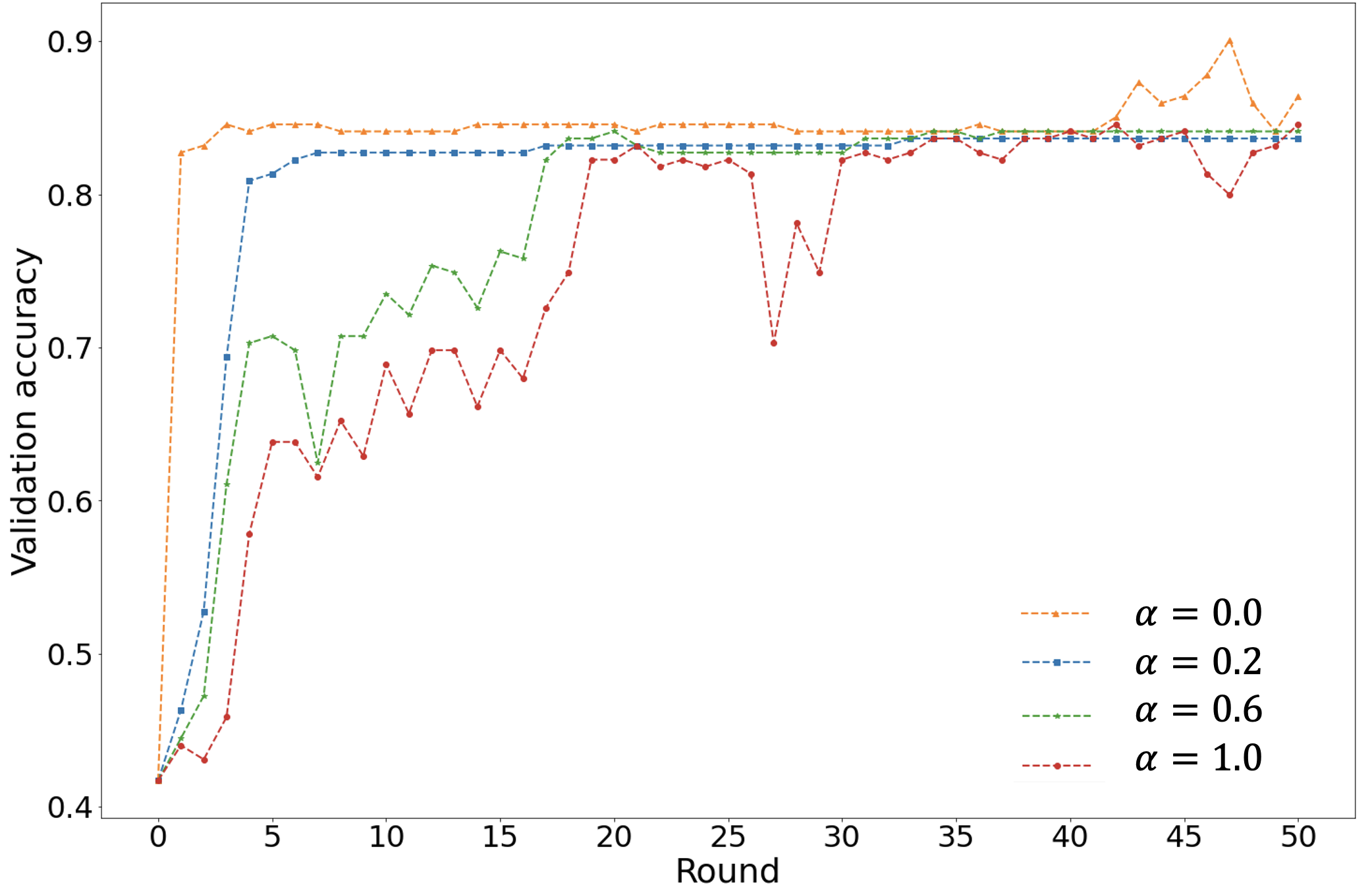}
            \caption{10 clients}
    \end{subfigure}%
    \begin{subfigure}[b]{0.33\textwidth}
            \centering
            \includegraphics[width=\textwidth]{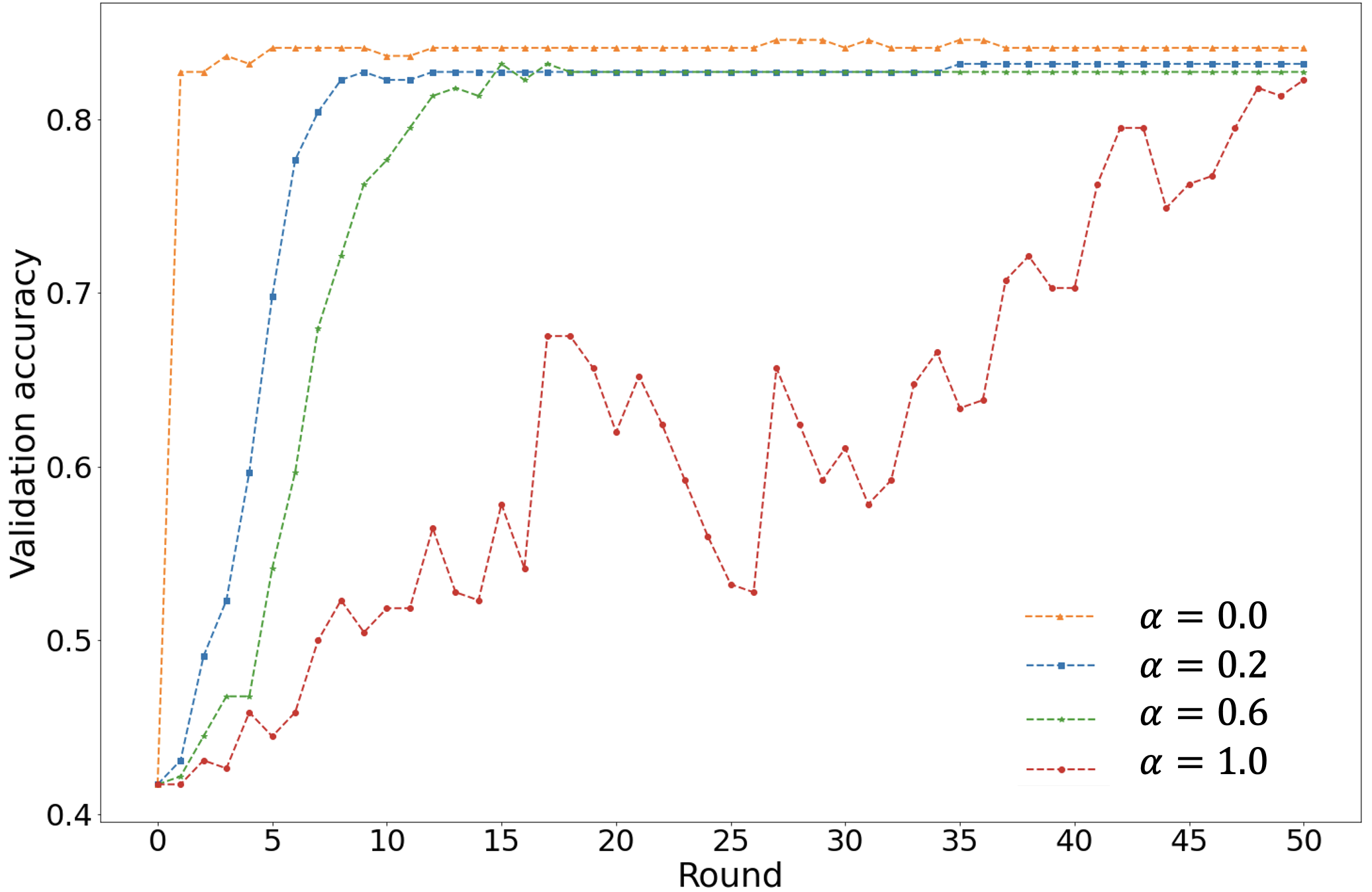}
            \caption{20 clients}
    \end{subfigure}
     \begin{subfigure}[b]{0.33\textwidth}
            \centering
            \includegraphics[width=\textwidth]{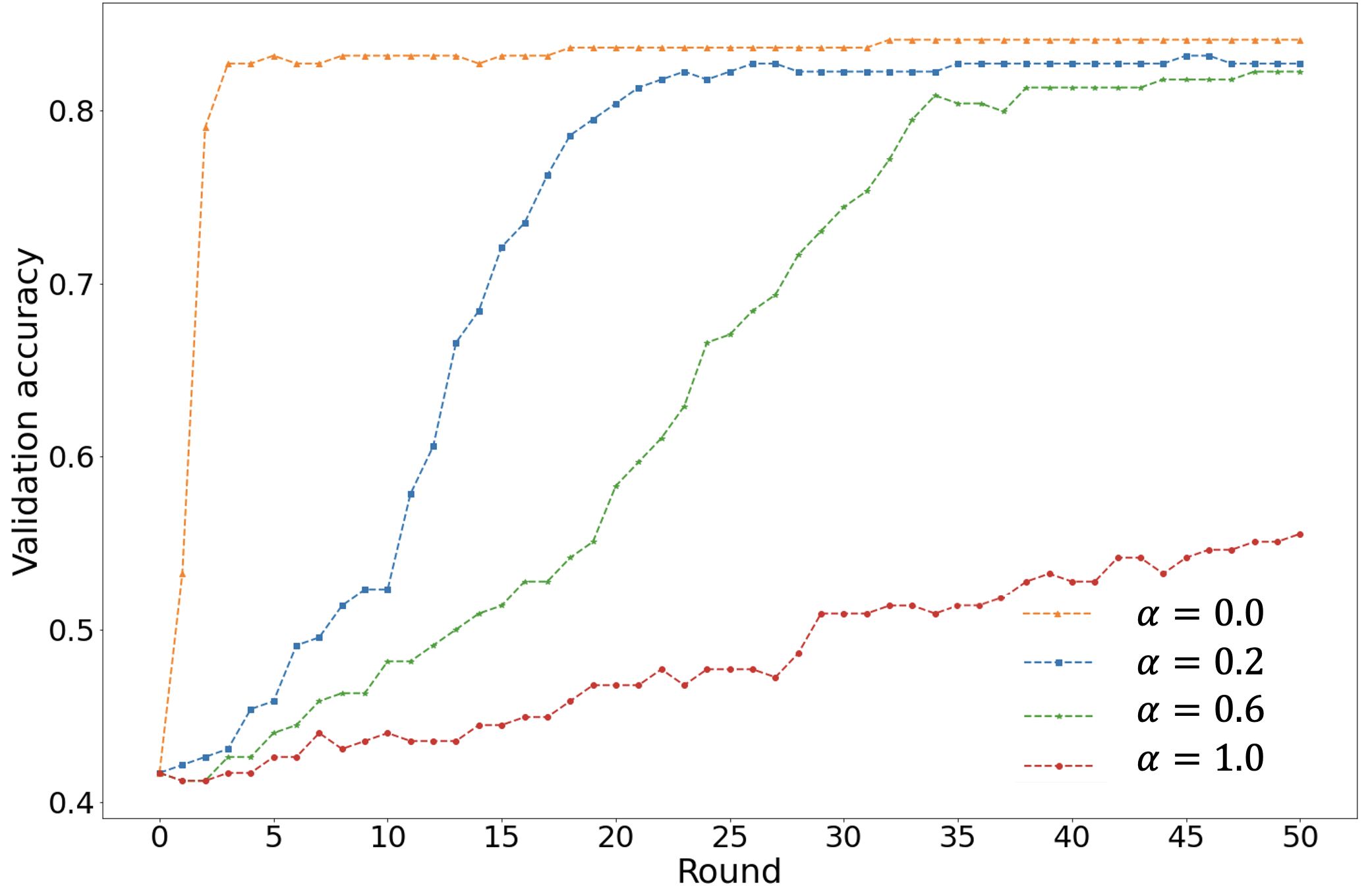}
            \caption{50 clients}
    \end{subfigure}
    \caption{Performance evaluation for the different levels of data heterogeneity.}
    \label{level}
\end{figure*}

To compare the performance of the FedPB, we conducted centralized learning with the same task model, by training the model on the entire dataset. Its performance was evaluated at the end of each round based on the test set. The early stopping was employed to monitor the variance in validation loss with a patience value of 10, thus automatically stopping the training when there appeared to be no decrease of the validation loss over the last 10 epochs. We applied Adam,with a learning rate of 0.0001 and a batch size of 16. Standalone learning refers to a client training a local task model without sharing knowledge with others. We evaluated standalone learning by training a local task model on each client's local dataset, respectively, with the same hyperparameters and training setting as in centralization learning. 

We first studied a 10-client scenario with independent and identically distributed (IID) local data in the FedPB. In this case, the training set was evenly divided into 10 subsets with the same number of phishing and legitimate emails. Then, for each round, a subset of three clients was randomly selected to conduct the local model training based on their local data. The local model training adopted the Adam with a learning rate of 0.0001, a batch size of 16, and an epoch of one. We evaluated the global model's performance using the test set for a total of 50 rounds and compared its performance with the other methods.

Furthermore, for each method, we performed 10 individual experiments with different seeds. Fig. \ref{result} illustrates the average test accuracy over the last five rounds of training for each method. The result shows that the FedPB can outperform standalone learning of different clients and achieve a competing performance with centralized learning. The decrease in the detection performance is considered due to the information loss during the model aggregation.

With a continually increasing client number, the performance of the system can greatly degrade due to more challenging aggregation with more local updates involved \cite{thapa}. To study the systems' robustness to different client numbers, we applied a client number $K\in\{10,20,50\}$ respectively to train the model. Moreover, to ensure the same amount of data was used in each case and the only variable was the client number, we randomly selected $K_{selected}\in\{3,6,15\}$ clients accordingly for each round's model training. We performed model training with the IID setting for a total of 50 rounds and evaluated the global model using the test set every round. Fig. \ref{clientnum} illustrates the global model’s performance at each round with 10, 20, and 50 clients respectively. 

\subsubsection{Varying the Data Heterogeneity Level}

\label{heterogenity} 
In real-life applications, samples held by clients are usually skewed with different In real-life applications, samples held by clients are usually skewed with different distributions. Such heterogeneity of client data can result in the slow convergence of the global model, and the divergence of a training algorithm \cite{survey}. Let $\alpha$ be the heterogeneity level of a client’s data distribution in FL. $\alpha = \mid 2P_k - 1\mid$, where $P_k$ denotes the probability of phishing email samples in the client$k$'s local dataset, and $\alpha$ takes the absolute value of the computed result. For example, $\alpha = 1.0$ indicates that all client data  belong to a single label that can be either phishing or legitimate, and $\alpha = 0.6$ indicates that 80\% data belong to one class and the remaining 20\% data belong to the other class. All clients share the same $\alpha$ by randomly assigning different clients with either more phishing samples or legitimate samples. We evaluated the systems' performance when applying different data heterogeneity levels $\alpha \in\{0.0,0.2,0.6,1.0\}$. 

Fig. \ref{level} illustrates the global model’s performance in the 10, 20, and 50 clients scenarios when applying different heterogeneity levels respectively. Consequently, the results show that the FedPB can retain a test accuracy of 0.83 within the 50 rounds in most cases. However, when the data heterogeneity level is 1.0 (each client has only a single class's data) in the 50-client scenario, the model did not converge within the 50 learning rounds. In more changeling cases, we would need additional rounds to learn the model.

\section{Conclusions}
\label{conclusion}
The adaptability of an ML-based phishing email detection system is largely restricted by accessible training data. Unfortunately, recent years' escalating privacy concerns have rendered centralized learning less viable for handling email contents containing personal data. In this study, we proposed the Federated Phish Bowl (FedPB) to allow decentralized phishing email analysis and detection by leveraging the global word embedding and BiLSTM-based detection. We evaluated model performance by various settings of the client numbers and data heterogeneity levels, showing that FedPB can retain a competing detection accuracy of 0.83 with great robustness. 

Though FedPB does not require the sharing of clients' data, it can still encounter threats such as backdoor attacks \cite{backdoor}, information-stealing attacks \cite{deepleakage}, and so forth. For future work, we aim to study proper defense in FedPB against such threats. In addition, we will look at reducing the waiting time for the PS to receive clients' updates via asynchronous learning \cite{khan, lu}.

\bibliography{infocom}

\begin{thebibliography}{10}
\providecommand{\url}[1]{#1}
\csname url@samestyle\endcsname
\providecommand{\newblock}{\relax}
\providecommand{\bibinfo}[2]{#2}
\providecommand{\BIBentrySTDinterwordspacing}{\spaceskip=0pt\relax}
\providecommand{\BIBentryALTinterwordstretchfactor}{4}
\providecommand{\BIBentryALTinterwordspacing}{\spaceskip=\fontdimen2\font plus
\BIBentryALTinterwordstretchfactor\fontdimen3\font minus
  \fontdimen4\font\relax}
\providecommand{\BIBforeignlanguage}[2]{{%
\expandafter\ifx\csname l@#1\endcsname\relax
\typeout{** WARNING: IEEEtran.bst: No hyphenation pattern has been}%
\typeout{** loaded for the language `#1'. Using the pattern for}%
\typeout{** the default language instead.}%
\else
\language=\csname l@#1\endcsname
\fi
#2}}
\providecommand{\BIBdecl}{\relax}
\BIBdecl

\bibitem{f5}
D.~Warburton, ``Phishing attacks soar 220\% during covid-19 peak as
  cybercriminal opportunism intensifies,''
  \url{https://www.f5.com/labs/articles/threat-intelligence/2020-phishing-and-fraud-report},
  2020, accessed: 2022-03-30.

\bibitem{gutierrez}
C.~N. Gutierrez, T.~Kim, R.~D. Corte, and et~al., ``Learning from the ones that
  got away: Detecting new forms of phishing attacks,'' \emph{IEEE Transactions
  on Dependable and Secure Computing}, vol.~15, no.~6, pp. 988--1001, 2018.

\bibitem{unnithan}
H.~M., N.~Unnithan, V.~Ravi, and S.~Kp, ``Deep learning based phishing e-mail
  detection cen-deepspam,'' \emph{ACM Conference on Data and Application
  Security and Privacy}, 2018.

\bibitem{nguyen}
M.~Nguyen, T.~Nguyen, and T.~H. Nguyen, ``A deep learning model with
  hierarchical lstms and supervised attention for anti-phishing,'' 2018.

\bibitem{smadi}
S.~Smadi, N.~Aslam, and L.~Zhang, ``Detection of online phishing email using
  dynamic evolving neural network based on reinforcement learning,''
  \emph{Decision Support Systems}, vol. 107, pp. 88--102, 2018.

\bibitem{sahingoz}
O.~K. Sahingoz, E.~Buber, O.~Demir, and B.~Diri, ``Machine learning based
  phishing detection from urls,'' \emph{Expert Systems with Applications}, vol.
  117, pp. 345--357, 2019.

\bibitem{zhang}
Y.~Fang, C.~Zhang, C.~Huang, and et~al., ``Phishing email detection using
  improved rcnn model with multilevel vectors and attention mechanism,''
  \emph{IEEE Access}, vol.~7, pp. 56\,329--56\,340, 2019.

\bibitem{alhogail}
A.~Alhogail and A.~aah Alsabih, ``Applying machine learning and natural
  language processing to detect phishing email,'' \emph{Computers and
  Security}, vol. 110, p. 102414, 2021.

\bibitem{thapa}
C.~Thapa, J.~W. Tang, A.~Abuadbba, and et~al., ``Evaluation of federated
  learning in phishing email detection,'' 2021.

\bibitem{dl}
Y.~LeCun, Y.~Bengio, and G.~Hinton, ``{Deep learning},'' \emph{Nature}, vol.
  521, no. 7553, pp. 436--444, 2015.

\bibitem{365}
Microsoft, ``Anti-phishing policies in microsoft 365,''
  \url{https://docs.microsoft.com/en-us/microsoft-365/security/office-365-security/set-up-anti-phishing-policies?view=o365-worldwide},
  2021, accessed: 2022-03-31.

\bibitem{bs}
L.~Richardson, ``Beautiful soup documentation,'' 2007.

\bibitem{fl}
J.~Konečný, H.~B. McMahan, F.~X. Yu, and et~al., ``Federated learning:
  Strategies for improving communication efficiency,'' in \emph{NIPS Workshop
  on Private Multi-Party Machine Learning}, 2016.

\bibitem{glove}
J.~Pennington, R.~Socher, and C.~D. Manning, ``Glove: Global vectors for word
  representation,'' in \emph{Empirical Methods in Natural Language Processing
  (EMNLP)}, 2014, pp. 1532--1543.

\bibitem{bidirectional}
M.~Schuster and K.~K. Paliwal, ``Bidirectional recurrent neural networks,''
  \emph{{IEEE} Trans. Signal Process.}, vol.~45, no.~11, pp. 2673--2681, 1997.

\bibitem{enron}
B.~Klimt and Y.~Yang, ``The enron corpus: A new dataset for email
  classification research,'' in \emph{ECML}, 2004.

\bibitem{survey}
Y.~Sun, H.~Ochiai, and H.~Esaki, ``Decentralized deep learning for multi-access
  edge computing: A survey on communication efficiency and trustworthiness,''
  \emph{IEEE Transactions on Artificial Intelligence}, 2021.

\bibitem{backdoor}
E.~Bagdasaryan, A.~Veit, Y.~Hua, D.~Estrin, and V.~Shmatikov, ``How to backdoor
  federated learning,'' in \emph{AISTATS}, 2020.

\bibitem{deepleakage}
L.~Zhu, Z.~Liu, and S.~Han, ``Deep leakage from gradients,'' in \emph{NeurIPS},
  2019.

\bibitem{khan}
L.~U. Khan, S.~R. Pandey, N.~H. Tran, W.~Saad, and et~al., ``Federated learning
  for edge networks: Resource optimization and incentive mechanism,''
  \emph{IEEE Communications Magazine}, vol.~58, no.~10, pp. 88--93, 2020.

\bibitem{lu}
Y.~Lu, X.~Huang, Y.~Dai, and et~al., ``Differentially private asynchronous
  federated learning for mobile edge computing in urban informatics,''
  \emph{{IEEE} Trans. Ind. Informatics}, vol.~16, no.~3, pp. 2134--2143, 2020.

\end{thebibliography}
\bibliographystyle{IEEEtran}

\end{document}